\documentstyle[epsfig,12pt]{article}
\begin{document}

\begin{center}
{\bf A formula for the number of days between the winter solstice and the latest sunrise} \\
\vspace{.07in}
Matthias Reinsch\\
Department of Physics, University of California,\\
Berkeley, California 94720\\
{\tt mreinsch@socrates.berkeley.edu, mreinsch@earthlink.net}\\
March 10, 2003
\end{center} 

\vspace{.2in}

\begin{quote}
A simple closed-form expression for the number of days between the winter solstice and the latest sunrise is derived.  Formulas for the summer solstice and the sunset are derived as well.  These approximate formulas make it easy to see the functional dependence on the latitude of the observer.  An approximate expression for the difference in time of day between the latest sunrise and the sunrise on the solstice is also derived.  The formulas are not valid in the Arctic, Antarctic or tropics.
\end{quote}

\vspace{.2in}
\noindent
{\bf Introduction}
\vspace{.1in}

In this paper, we present a simple closed-form expression for the number of days between the winter solstice and the latest sunrise.  This formula is the result of a lengthy calculation which simplifies greatly in the end.  A few approximations are made in the derivation of the result, but it agrees well with actual data for sunrise times.

Let us begin by recalling the reasons why the latest sunrise does not occur on the shortest day of the year.  There are two effects that contribute to this phenomenon.  The dominant effect is that due to the inclination of the Earth's axis of rotation.  The other contribution is due to the eccentricity of the Earth's orbit.

We first discuss the effect due to the inclination of the Earth's axis of rotation, that is, we will begin by taking the Earth's orbit to be a perfect circle.  It is useful to work with the concept of the celestial sphere.  In the course of a year the sun's path is a great circle on the celestial sphere.  This great circle is inclined at an angle of $23.44^o$ relative to the celestial equator.  The sun moves along this great circle, the ecliptic, at a constant rate, because we are assuming that the Earth's orbit is a perfect circle.  Thus, the right ascension of the sun does not change at a constant rate. (The right ascension is the east/west coordinate on the celestial sphere that plays the same role as the longitude coordinate on the surface of the Earth.)  At a solstice, the right ascension of the sun changes the fastest because the ecliptic is closest to one of the poles there.  At an equinox, the right ascension changes the slowest because the ecliptic intersects the celestial equator at an angle.

Thus, if a picture is taken of the sky at 12:00 noon every day for a year, we see not only the sun's familiar north/south motion, but also an east/west motion.  The reason is that 24 hours is the length of an average solar day, but we have seen that during the year the sun's right ascension changes at a non-uniform rate.  The images taken of the sun would trace out a certain figure eight, called the analemma [1].  When drawing the numeral eight with a pen, we see that the pen moves in the same direction at the top and bottom, and in the direction opposite to this during both passages through the center.  This corresponds to what we said above about the solstices versus the equinoxes.

In particular, we are interested in the fact that the sun is moving in the east/west direction in this sequence of photos, at the winter solstice.  In the northern hemisphere, the day (in the sense of the amount of time that the sun is visible above the horizon) is shortest on the winter solstice, but because of the east/west motion of the sun, the times of both the sunrise and the sunset are still changing at this time of year (by the same amount, because their difference is minimal at this time).  Thus, the latest sunrise is not on the shortest day of the year.

Next, we include the eccentricity of the Earth's orbit into the discussion.  The path of the sun on the celestial sphere is not changed.  What is different is that the sun is no longer moving along this path at a constant rate.  Thus, the rate of change of the right ascension of the sun is changed.  The result is that the analemma discussed above becomes lop-sided, and this modifies the amount of east/west motion at the solstices.

\vspace{.2in}
\noindent
{\bf Statement of result}
\vspace{.1in}

Let $n$ denote the number of days between the winter solstice and the latest sunrise. Then at
latitude $\lambda$, assumed to be not too close to the equator or the poles as explained below, $n$ is given approximately by

\begin{equation}
\label{formula}
n = \left(\delta - \frac{2 \pi}{N_y} \cos \alpha \right) \frac{\sqrt{\cos(2 \alpha) + \cos(2 \lambda)}}
{\sqrt{2} \, \delta^2 \tan \alpha \, \sin \lambda } \, ,
\end{equation}
where
\begin{equation}
N_y = 365.2564
\end{equation}
is the number of ephemeris days per sidereal year [2], and
\begin{equation}
\alpha = 23.44^o
\end{equation}
is the angle between Earth's rotation axis and the normal to the ecliptic, and
\begin{equation}
\label{delta}
\delta = \frac{2 \pi \sqrt{1 - \epsilon^2} }{N_y (1 - \epsilon \cos \psi_o)^2}
\end{equation}
is the change in the angular position of the Earth about the Sun per ephemeris day at the winter solstice [3].  In this expression,
\begin{equation}
\epsilon = .01672
\end{equation}
is the eccentricity of the Earth's orbit, and
\begin{equation}
\psi_o = 1.92 \pi
\end{equation}
is the eccentric anomaly at the winter solstice.  It should be noted that this value for $\psi_o$ is approximate; its exact value does not affect the result for $n$ significantly.  The value of $\psi_o$ can be determined from the date of the perihelion if one assumes that the Earth's orbit is an ellipse.  In actuality there are perturbations to the orbit, and the dates of the perihelions vary quite a bit.  According to the Astronomical Applications Department of the U.S. Naval Observatory (http://aa.usno.navy.mil/data/docs/EarthSeasons.html), the dates of the perihelions in the years from 1993 to 1998 were Jan.\ 4, Jan.\ 2, Jan.\ 4, Jan.\ 4, Jan.\ 2 and  Jan.\ 4. 

We see that the expression for $n$ given in Eq.\ (\ref{formula}) becomes larger as we approach the tropics and that it goes to zero as we approach the Arctic or Antarctic Circle.  This is because the $\cos(2 \alpha) + \cos(2 \lambda)$ in the numerator goes to zero as we approach the Arctic or Antarctic Circle.  The formula is not valid for values of $\lambda$ that are north of the Arctic Circle or south of the Antarctic Circle because the sun does not rise and set on a daily basis in those regions.  The other dependence on the latitude $\lambda$ is the $\sin \lambda$ in the denominator.  This formula is not intended for use in the tropics, as explained later in this paper, so there is no problem with a divergence.

The conventions that we use regarding the sign of $\lambda$ are that positive values are used in the northern hemisphere and negative values are used in the southern hemisphere.  In the northern hemisphere, the values for $n$ are positive, indicating that the latest sunrise is after the winter solstice.  In the southern hemisphere, the values for $n$ are negative, indicating that the extremal (earliest) sunrise is before the December solstice.  The term ``December solstice'' can be used to avoid confusion in these considerations.

We have also derived formulas for the June solstice and the sunset.  The results may be summarized as follows.  For sunsets the value of $n$ is always minus the corresponding sunrise $n$ at the same location and for the same solstice.  For the June solstice sunrise, the formula for $n$ is minus the formula in Eq.\ (\ref{formula}), with the appropriate $\delta$ value used.  This $\delta$ value can be found from Eq.\ (\ref{delta}) using the value of the eccentric anomaly for the June solstice, which is approximately $\pi$ less than the value at the December solstice.  These statements regarding the different formulas for $n$ are made under the assumption that the angle of inclination $\alpha$ is not greater than $90^o$.  This is true for most planets in our solar system.

\vspace{.2in}
\noindent
{\bf Examples}
\vspace{.1in}

Table I compares the approximation given in Eq.\ (\ref{formula}) with tabulated values from the Astronomical Applications Department of the U.S. Naval Observatory \\
(http://aa.usno.navy.mil/data/docs/RS\_OneYear.html) for the solstice of December 2001.  In tables where times are rounded to the nearest minute, there will be a range of dates that have the latest sunrise time.  The corresponding range of $n$ values is given in the column labeled ``actual value of $n$.''  We do not expect our approximations regarding atmospheric refraction and the size of the solar disk to be good near the Arctic Circle.  Near the tropics, $n$ is so large that the Taylor expansion to second order is not sufficient.

\vspace{.1in}

\newpage

\begin{tabbing}
\indent \= location (city, state) 
\qquad \= latitude (degrees) 
\qquad \= approximate $n$ 
\qquad \=\kill
\>  \> $\; \;  {\rm Table \;I}$\>  \>  \\ [6pt]
\> location (city, state) \> latitude (degrees) \> approximate $n$ \> actual value of $n$ \\ [6pt]
\> Fairbanks, Alaska \> 64.82 \> 2.4 \> 2 - 5 \\
\> Anchorage, Alaska \> 61.22 \> 4.5 \> 1 - 8 \\
\> San Francisco, California \> 37.77 \> 16.2 \> 14 - 16 \\
\> Key West, Florida \> 24.55 \> 28.6 \> 16 - 31
\end{tabbing}

\vspace{.1in}

Another comparison of the approximation given in Eq.\ (\ref{formula}) with actual sunrises is given in Fig. 1.  The comparison is made for a range of latitudes, from the Tropic of Cancer to the Arctic Circle.

\vspace{.2in}
\noindent
{\bf Derivation of Eq.\ (\ref{formula}) }
\vspace{.1in}

Since the derivation of Eq.\ (\ref{formula}) is a lengthy calculation, we give only a summary here.  The analysis begins by describing the motion of the Earth in a heliocentric ecliptic coordinate system, as described in Ref.[3].  We use the angle $\theta$ to denote the angular position of the Earth relative to the Sun.  This angle also gives the position of the sun in a geocentric ecliptic coordinate system.  We define $\theta$ to be zero at the vernal equinox, as is standard.  Next, we transform to a topocentric equatorial coordinate system, and then to a horizontal coordinate system.  This final coordinate system is the one an observer at a given point on the Earth would use, and an angle $h$ is used to measure the elevation of objects above the horizon, with $h = 0^o$ at the horizon and $h = 90^o$ at zenith.  The result for the sun is
\begin{equation}
\sin h = \cos\lambda \cos\theta\cos\phi+\cos\alpha \cos\lambda\sin\theta\sin\phi
+\sin\alpha\sin\lambda \sin\theta \, .
\label{sinofh}
\end{equation}
This equation gives $\sin h$ at any instant in time.  The angle $\phi$ is an angle that measures the rotation of the Earth about its own axis.  The time dependence of $\sin h$ is due to the time dependences of $\theta$ and $\phi$.  The angle $\phi$ evolves in time at a constant rate, for our purposes, increasing by $2\pi$ every sidereal day.  The angle $\theta$ increases by $2\pi$ every sidereal year, but not at a constant rate, due to the eccentricity of the Earth's orbit.

If we ignore refraction in the Earth's atmosphere and the finite size of the solar disk [4], we could find the times of sunrise and sunset by setting $\sin h$ equal to zero and solving the transcendental equation that results for the time $t$.  The errors made in this way should be more or less constant over a small part of a year, and we hope that they do not affect our search for the latest sunrise too much.  The next approximation is to use the same value for $\theta$ throughout any given day.  This makes it possible to solve for the times of sunrise and sunset using elementary functions.  Again, it should be the case that these errors do not vary significantly over a small part of the year and thus do not affect the day with the latest sunrise excessively.  Now that we have a formula for the time of the sunrise on any given day, we make a second-order Taylor series expansion about the winter solstice.  This is where the calculation gets complicated, but with enough work the coefficients can be simplified.  Finally, the extremum of a quadratic function can be found easily, and the result is Eq.\ (\ref{formula}).  More details about this calculation are given in Appendix~A.\  An approximate expression for the difference in time of day between the latest sunrise and the sunrise on the solstice is also derived there.  

We note that Eq.\ (\ref{formula}) does not behave well when $\alpha$ goes to zero.  Apart from situations where $\alpha$ is very small, this result should be applicable to other planets.  As $\alpha$ goes to zero, the quadratic term in the Taylor expansion discussed above vanishes, so a higher-order expansion would be necessary.  

At the beginning of this paper, we stated that the results were intended for use at latitudes that were not too close to the equator or the poles.  We can now clarify these limits.  Using terminology appropriate for the northern hemisphere, we can say that the formula should not be used for latitudes north of the Arctic Circle because the sun does not rise and set on a daily basis there throughout the year.  Such daily risings and settings were assumed in the calculation.  From the point of view of mathematics, the formula gives imaginary numbers north of the Arctic Circle.  The Tropic of Cancer is more of a soft limit.  The value for $n$ is so large at the Tropic of Cancer, that the second-order Taylor expansion is no longer sufficient, and the results become less accurate the further south one goes (see Fig.~1).

\vspace{.2in}
\noindent
{\bf Classroom activities and projects}
\vspace{.1in}

The materials for our first activity are a globe and a measuring tape made of cloth or paper.  The idea is to get a better understanding of the fact that the right ascension of the sun changes at different rates at different times of the year.  The lines of longitude and latitude on the globe form the same type of coordinate system as the one used on the celestial sphere.  The purpose of the globe in this demonstration is to provide a sphere with such a coordinate system, and we will think of this sphere as representing the celestial sphere.  Place the beginning of the measuring tape at the point on the globe that has longitude and latitude both equal to zero.  Then wrap the tape around the globe so that it traces out a great circle that is inclined relative to the equator.  An angle of inclination of about 23 degrees would correspond to the inclination of the Earth's axis as discussed in the Introduction, and the measuring tape would then represent the ecliptic.  Now choose a sequence of evenly spaced points along the measuring tape, and find the right ascension of each point (the longitude on the globe).  Plotting these values on graph paper will show that the right ascension changes the slowest when the ecliptic crosses the equator (at the equinoxes) and fastest when the ecliptic is furthest from the equator (at the solstices).  The effect is easier to see if a larger angle of inclination is used.

An extension of these ideas can be used to create a plot of the analemma.  Make a table of the right ascension and declination (longitude and latitude) of 36 evenly spaced points on the ecliptic, starting at the beginning of the measuring tape and going west around globe.  The declination column of the table will also be called the $Y$ column.  Next, we will create a third column of the table, which we will call the $X$ column.  The values for this column are obtained by taking the values in the right ascension column and subtracting an integer multiple of 10 degrees.  The integer is 0 for the first row, 1 for the second row, 2 for the third row, and so on.  The result is that the $X$ and $Y$ columns give the position of the sun relative to its average position at 12:00 noon (see the discussion in the Introduction).  Now create a plot with $X$ and $Y$ axes.  The data points should trace out a figure eight.  The real analemma is actually lop-sided, due to the eccentricity of the Earth's orbit.  Some globes have a picture of the analemma, often placed somewhere in the Pacific Ocean where there is space for it.

For our next activity, work out the value of $n$ at your latitude using Eq.\ (\ref{formula}).   Compare the results with values in published tables, such as those found in tide books.  You can also use one of the the web pages of the Astronomical Applications Department of the U.S. Naval Observatory, http://aa.usno.navy.mil/data/docs/RS\_OneYear.html ,  to generate a table of sunrise and sunset times for your location.  Next, imagine living on Mars at the same latitude.  What would the value of $n$ be there?  It will be necessary to look up various orbital parameters for Mars in order to calculate this, and the length of the Martian day will be needed.

The materials for our last activity are a watch and a piece of chalk.  In the Introduction we mentioned a sequence of photographs taken of the sky every day at 12:00 noon.  The purpose of these photographs was to record the location of the sun at the same time on different days.  We can carry out similar measurements by recording the location of the shadow of the top of a flag pole or other convenient object at the same time on a sequence of days.  If class is not in session at 12:00 noon, another time of day can be used.  The greatest motion of the shadow from one day to the next will be seen in March and September, near the equinoxes.  At these times, the component of the shadow's movement in the North/South direction will be about 0.4 degrees per day.  Shadows cast by the sun are a bit fuzzy because of the nonzero size of the solar disk (about half a degree).  Therefore, the motion can be seen more clearly by waiting for several days to pass.  The East/West motion in which we are interested in this paper is large near both the solstices and the equinoxes.  Its maximal value is on the order of 0.1 degrees per day.  At the equinoxes the motion will be in the direction opposite to the motion at the solstices, as discussed in the Introduction.  If the shadow falls on an area where there are tiles, an alternative to using chalk marks on the ground would be to use the coordinate system provided by the tiles to record the position of the shadow.  A larger project would be to record the location of the shadow once a week or so, for an entire semester or even a year.

\vspace{.2in}
\noindent
{\bf Acknowledgements}
\vspace{.1in}

The author wishes to thank Jim Morehead for many useful discussions.

\newpage

\vspace{.2in}
\noindent
{\bf Appendix A}
\vspace{.1in}

This appendix gives a more detailed description of the derivation of Eq.~(\ref{formula}).  The angle $\phi$ in Eq.~(\ref{sinofh}) measures the rotation of the Earth about its own axis.  The value of $\phi$ as a function of time is
\begin{equation}
\phi(t) = \left( \frac{2\pi}{T_{sidereal}} \right) t \, ,
\end{equation}
where $T_{sidereal} \approx 86164.09$ seconds is the time it takes for the Earth to rotate once about its axis relative to the fixed stars.  For our calculation, it is useful also to define
\begin{equation}
\tilde\phi(t) = \left( \frac{2\pi}{T_{ephemeris}} \right) t \, ,
\end{equation}
with $T_{ephemeris} = 86400$ seconds $= 24$ hours, and the difference
\begin{equation}
\Delta(t) = \phi(t) - \tilde\phi(t) \, .
\end{equation}
Substituting $\tilde\phi + \Delta$ in for $\phi$ in Eq.~(\ref{sinofh}) and using some trigonometric identities gives
\begin{equation}
\sin h = A \cos(\tilde\phi + \zeta) \cos\lambda + \sin\alpha \sin\lambda \sin\theta \, ,
\end{equation}
where
$A$ and $\zeta$ are defined to be
\begin{equation}
A = \sqrt{(\cos\alpha \sin\theta \sin\Delta + \cos\theta \cos\Delta)^2 +
(\cos\alpha \sin\theta \cos\Delta - \cos\theta \sin\Delta)^2} \, ,
\end{equation}
and
\begin{equation}
\zeta = -\tan^{-1}\left(
\frac{\cos\alpha \sin\theta \cos\Delta - \cos\theta \sin\Delta}
{\cos\alpha \sin\theta \sin\Delta + \cos\theta \cos\Delta}\right) \, .
\end{equation}
The argument of the $\tan^{-1}$ function on the right-hand side of the equation for $\zeta$ stays small throughout the course of the year, so there is never a question about which branch of the $\tan^{-1}$ function to use.  One way to see this is to first imagine replacing $\cos\alpha$ with $1$, whereupon the numerator becomes $\sin(\theta - \Delta)$ and the denominator becomes $\cos(\theta - \Delta)$.  Both $\theta$ and $\Delta$ increase monotonically from $0$ to $2\pi$ over the course of the year, so the numerator stays close to $0$ and the denominator stays close to $1$.  In actuality, $\cos\alpha$ is not $1$, but the functions still stay close to $0$ and $1$, respectively.

Over the course of the year, the values of $\phi$ modulo $2\pi$ at sunrise and sunset vary greatly, because $\phi$ measures the rotation of the Earth relative to the fixed stars.  In contrast, the values of $\tilde\phi$ modulo $2\pi$ at sunrise and sunset do not vary by such large amounts, because $\tilde\phi$ was defined to be an angle that increases by $2\pi$ every 24 hours (mean solar day in the year 1900).  This is the reason for transforming from $\phi$ to $\tilde\phi$.  If we now take the values of $\theta$ and $\Delta$ to be constant on any given day, we can solve for the times of day of sunrise and sunset.  Expressed in terms of $\tilde\phi$ (modulo 24 hours, so $\tilde\phi$ is in the range from $-\pi$ to $\pi$) these are
\begin{equation}
\tilde\phi = - \zeta \pm \cos^{-1}\left( \frac{-\sin\alpha \tan\lambda \sin\theta}
{A} \right),
\label{riseset}
\end{equation}
where the upper sign is for the sunset and the lower sign is for the sunrise.  Note that at the vernal equinox, $\tilde\phi$ and the time $t$ are zero, and the sun is crossing the meridian at the location of the observer.  Thus, $\tilde\phi$ is zero near noon, and sunrise times have a negative value for $\tilde\phi$, while sunset times have a positive value for $\tilde\phi$.

The lower sign in Eq.~(\ref{riseset}) gives an explicit formula for the time of day of the sunrise on each day of the year (expressed in terms of a $\tilde\phi$ value).  We are interested in finding the extremum near the winter solstice.  We use an integer $k$ to index the days relative to the winter solstice [$k = 0$ on the winter solstice, $k = 1$ on the next day, etc.].  Then at some time on day $k$ the value of $\theta$ is (to order $k^2$)
\begin{equation}
\theta(k) = \frac{3}{2}\pi + C_1 k + \frac{1}{2} C_2 \, k^2 \, ,
\end{equation}
where $C_1$ is $d\theta/dt$ evaluated at the winter solstice, times 24 hours, and $C_2$ is
$d^2\theta/dt^2$ evaluated at the winter solstice, times the square of 24 hours.  A similar expansion can be written out for $\Delta(k)$, but it does not contain a $k^2$ term because $\Delta$ is a linear function of time.  When these expressions are substituted into the formula for the time of day of the sunrise, the result is a function of $k$ that is accurate to order $k^2$.  By calculating the first and second derivatives of this function with respect to $k$, we can calculate a quadratic approximation for the function, and from this find the value of
$k$ at the extremum.  We use $n$ to denote this extremal value of $k$, and the result is given in Eq.~(\ref{formula}).  There is actually a small correction term involving the $C_2$ coefficient mentioned above, but this term is smaller than the result stated in Eq.~(\ref{formula}) by a factor of $\epsilon$, so it can be neglected.

The quadratic approximation can also be used to calculate another quantity of interest.  We let $\Delta t$ denote the difference in time of day between the latest sunrise and the sunrise on the winter solstice.  This time difference can be calculated by substituting the value of $n$ back into the quadratic approximation for the time of day of the sunrise.  The result can be expressed as
\begin{equation}
\Delta t = \frac{T_{ephemeris}}{2} \,
\left( \frac{\delta}{2 \pi \cos \alpha} - \frac{1}{N_y} \right) \, n \, .
\end{equation}
Evaluating the constants, we get that $\Delta t$ is approximately $15$ seconds times $n$.  It is interesting that the constant of proportionality is independent of the latitude.  Sample numbers for San Francisco are $\Delta t = 245$ seconds and $n$ = 16, with a ratio approximately equal to $15$ seconds.

\vspace{.2in}
\noindent
{\bf References}
\vspace{.1in}

[1] di Cicco, Dennis, ``Astro Imaging –- Photographing the Analemma,'' Sky \& Telescope, March 2000, p. 135-140.

[2] Ottewell, Guy, {\em The Astronomical Companion}, published at the Department of Physics, Furman University, Greenville, South Carolina, 1979.

[3] Montenbruck, Oliver, {\em Practical ephemeris calculations}, Springer-Verlag, Berlin, New York, 1989.

[4] Meeus, Jean, {\em Astronomical algorithms, 2nd ed., 2nd English ed.}, Willmann-Bell, Richmond, Va., 1998.

\newpage

\epsfig{figure=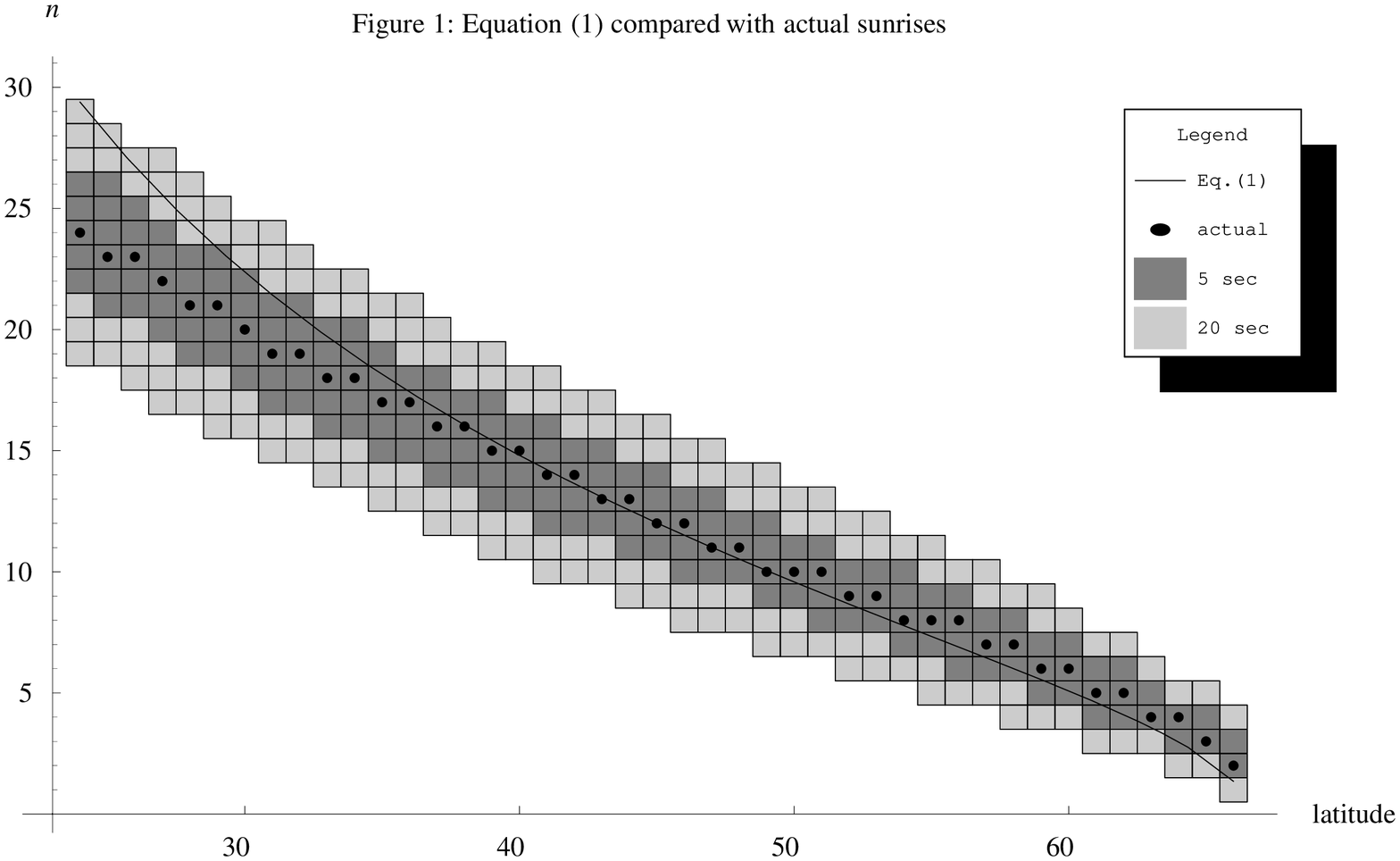,height=4.0in}

\vspace{.15 in}

The result presented in this paper is compared with actual sunrises.  The curve is a plot of $n$ (the number of days between the winter solstice and the latest sunrise) as a function of the latitude $\lambda$, as given by Eq.~(1).  The dots are the actual values of $n$, plotted for latitudes that are integer numbers of degrees.  Gray squares are also plotted for latitudes that are integer numbers of degrees.  A darker gray square at a value of $n$ and $\lambda$ means that the time of the sunrise on the $n$-th day after the solstice is within 5 seconds of the latest sunrise at that latitude.  The lighter gray square is for a value of 20 seconds.  The ``actual'' times for sunrises were computed as discussed in Ref.~[4], using the standard definition of sunrise to be the time when the center of the sun is 50 arc minutes below the horizon.  The 50 arc minutes take into account the size of the solar disk and the effects of atmospheric refraction.

\end{document}